\documentclass[10pt,a4paper]{article}
\usepackage{mathtools}
\usepackage{mathrsfs}
\usepackage{amsmath}
\usepackage[T1]{fontenc}
\usepackage[utf8]{inputenc}
\usepackage{amssymb}
\usepackage{multicol}
\usepackage{bbm}
\usepackage{bm}
\usepackage[usenames,dvipsnames]{color}
\usepackage[draft]{todonotes}
\usepackage[a4paper,top=2cm,bottom=2cm,left=2cm,right=2cm,bindingoffset=5mm]{geometry}
\usepackage{booktabs} 
\usepackage{tikz}
\usepackage{cite}

\usepackage{enumitem}

\usepackage{booktabs} 
\usepackage{multirow}
\usepackage{bbm}
\usepackage{hyperref}
\usepackage{booktabs}
\usepackage{array}
\usepackage{stmaryrd}
\newcolumntype{?}{!{\vrule width 1pt}}

\newcommand{\red}[1]{{\color{red}#1}}
\newcommand{\blue}[1]{{\color{blue}#1}}

\newcommand{\q}{q}

\newcommand{\mP}{\mathbb{P}}
\newcommand{\mT}{\mathbb{T}}

\def\spvecA#1;{\if;#1;\else #1\cr \expandafter \spvecA \fi}

\newcommand*{\colorboxed}{}
\def\colorboxed#1#{%
  \colorboxedAux{#1}%
}
\newcommand*{\colorboxedAux}[3]{%
  \begingroup
    \colorlet{cb@saved}{.}%
    \color#1{#2}%
    \boxed{%
      \color{cb@saved}%
      #3%
    }%
  \endgroup
}

\usepackage{hyperref}

\hypersetup{
colorlinks=true,         
linkcolor=blue,          
citecolor=red,        
urlcolor=red            
}

\title{\vspace{-1.5cm}\textbf{Physical Constraints on Quantum Deformations of Spacetime Symmetries}}

\author{Flavio~Mercati\footnote{\href{mailto:flavio.mercati@gmail.com}{flavio.mercati@gmail.com}} ~and Matteo~Sergola\footnote{\href{mailto:matteo.sergola@gmail.com}{matteo.sergola@gmail.com}}
\vspace{12pt} \\
\it Dipartimento di Fisica, Universit{\`a} di Roma ``La Sapienza'',\\
\it P.le A. Moro 2,00185 Roma, Italy.
}

\begin{document}

\maketitle

\vspace{-1cm}

\begin{abstract}
In this work we study the deformations into Lie bialgebras of the three relativistic Lie algebras: de Sitter, Anti-de Sitter and Poincar\'e, which describe the symmetries of the three maximally symmetric spacetimes. These algebras represent the centrepiece of the kinematics of special relativity (and its analogue in (Anti-)de Sitter spacetime), and provide the simplest framework to build physical models in which inertial observers are equivalent. Such a property can be expected to be preserved by Quantum Gravity, a theory which should build a length/energy scale into the microscopic structure of spacetime. Quantum groups, and their infinitesimal version `Lie bialgebras', allow to encode such a scale into a noncommutativity of the algebra of functions over the group (and over spacetime, when the group acts on a homogeneous space). In 2+1 dimensions we have evidence that the vacuum state of  Quantum Gravity is one such `noncommutative spacetime' whose symmetries are described by a Lie bialgebra. It is then of great interest to study the possible Lie bialgebra deformations of the relativistic Lie algebras. In this paper, we develop a characterization of such deformations in 2, 3 and 4 spacetime dimensions motivated by physical requirements based on dimensional analysis, on various degrees of `manifest isotropy' (which implies that certain symmetries, \emph{i.e.} Lorentz transformations or rotations, are `more classical'), and on discrete symmetries like P and T.
On top of a series of new results in 3 and 4 dimensions, we find a no-go theorem for the Lie bialgebras in 4 dimensions, which singles out the well-known `$\kappa$-deformation' as the only one that depends on the first power of the Planck length, or, alternatively, that possesses `manifest' spatial isotropy.
\end{abstract}

Keywords: Lie bialgebras; Hopf algebras; Quantum groups; Deformations of relativistic kinematics; Noncommutative spacetimes; Quantum gravity; Poincar\'e group; (Anti-) de Sitter group.

\section{Introduction}

In the last decades the possibility of \textit{deforming} (rather than breaking) the relativistic symmetries of empty space at the Planck scale ($E_p\sim L_p^{-1} \sim 10^{19} GeV $\footnote{We use units in which $\hbar = c =1$}) has received a considerable amount of attention in the Quantum Gravity community. One reason for this comes from 2+1 dimensional Quantum Gravity, which, because it lacks local propagating degrees of freedom (gravitons), can be quantized with topological QFT methods. Coupling this theory to matter and integrating away the gravitational degrees of freedom, one ends up with a nonlocal effective theory~\cite{matschull,freidel}. The spacetime symmetries that leave this theory invariant are not described by the Poincar\'e group. In its place, one finds a  Hopf algebra (or ``quantum group'' \cite{presley, majid22, majid}), which is a deformation of a Lie group into a noncommutative object, that depends on a scale with the dimensions of an energy (the Planck scale).
Further evidence supporting the emergence of noncommutative-geometric structures in quantum gravity is provided by String Theory~\cite{SeibergWitten}, in which the B-field can take an expectation value such that the string dynamics is effectively described by a field theory on a noncommutative spacetime. The spacetime symmetries of such field theories are described by a Hopf algebra~\cite{Watts2000}.\footnote{Hopf algebras emerged in several other contexts in physics, \emph{e.g.} (quantum) integrable models~\cite{IntegrableSigmaModels}, perturbative QCD~\cite{HopfAlgebras_n_PerturbativeQCD} and AdS/CFT \cite{HopfAlgebras_n_StringTheory} to cite some that are closer to the concerns of the present paper.}

The energy scale $E_p$ is introduced, in the Hopf-algebraic context, in a frame-independent way: there is still a 10-dimensional group of symmetries that allows to connect different inertial observers, but the transformation laws are deformed in an $E_p$-dependent way. There is no breaking of Lorentz invariance, just a deformation of Special Relativity into a theory with two invariant scales ($c$ and $E_p$), sometimes dubbed \emph{Doubly Special Relativity}~\cite{secondo, oriti}. In 3+1 dimensions we cannot reproduce the results obtained in 2+1D Quantum Gravity~\cite{matschull,freidel} because we lack the same level of understanding of the quantum theory of gravity, but we can conjecture that a similar mechanism is at work, and the `ground state' of quantum gravity is not Minkowski spacetime, but rather a quantum homogeneous space whose symmetries are described by a Hopf algebra.
The study of the possible Hopf algebra deformations of the Poincar\'e group in 3+1 dimensions then acquires great interest.

In the present work, we will be concerned with the infinitesimal version of quantum groups: Lie bialgebras. These play, for quantum groups, the same role that Lie algebras play for Lie groups: they describe the structure of the group in an  infinitesimal neighborhood of the identity. A quantum group is essentially a Lie group with an additional layer of structure that allows to make the algebra of functions on the group noncommutative. In Lie bialgebras this noncommutative structure is linearized. So a Lie bialgebra can be seen as two Lie algebras, one playing the traditional role of commutation relations between generators, and the other playing a novel role associated with the noncommutative structure. These two Lie algebras have to satisfy certain consistency conditions, namely, that the dual map to the Lie bracket of one algebra has to be a cocycle with respect to the Lie bracket of the other algebra.
A theorem due to Drinfel'd~\cite{Drinfeld} established that, modulo global issues, a quantum group can be completely described in terms of its Lie bialgebra.\footnote{Specifically, Drinfel'd theorem states that Lie bialgebras are in one-to-one correspondence with Poisson-Lie structures on the (simply connected) group obtained via the exponential map, and the quantization of the unique Poisson-Lie structure associated to a Lie bialgebra will be the quantum group that corresponds to it~\cite{Drinfeld,Semenov,Bidegain}.}

Our goal in this paper is to characterize the possible deformations of the three maximally symmetric algebras of relativistic symmetries: $\mathfrak{iso}(d-1,1)$, $\mathfrak{so}(d,1)$ and $\mathfrak{so}(d-1,2)$ in $d=2$, $3$ and $4$ spacetime dimensions. To this extent in what follows we shall use the word \textit{classification} as to a listing of all possible deformed Lie bialgebras after the constraints of covariance principles and dimensional analysis.  Strictly speaking, a true classification of Lie bialgebras would involve identifying the equivalence classes under automorphisms of the Lie algebras. In this sense we do not present classification results. What we do is to identify the Lie bialgebras that satisfy certain covariance principles and certain physically-motivated dimensional analysis requests. These constraints greatly reduce the number of `physically interesting' cases, and within this smaller set, the discussion of automorphisms becomes much simpler. In a commutative spacetime $\mathfrak{iso}(d-1,1)$, $\mathfrak{so}(d,1)$ and $\mathfrak{so}(d-1,2)$ describe the symmetry groups of Minkowski, de Sitter and Anti-de Sitter spaces, respectively. We are then considering the possible symmetries of quantum homogeneous spacetimes, which are natural candidates for the vacuum state of quantum gravity.

In \ref{math} we will present a brief introduction of Lie bialgebras and their mathematics. In Section \ref{bia} we will describe our approach to classifying the Lie bialgebras which can be constructed from the Poincar\'e and (A)dS algebras in dimensions lower or equal to 3+1. In Section \ref{sol} we will list all the deformations we are able to find, in a systematic manner. In Section \ref{conc}  present our conclusions with future perspectives.

\section{Mathematical Tools: Lie bialgebra Basics}\label{math}

A Lie bialgebra\footnote{For a complete treatment we refer the interested reader to \cite{presley,104,majid}.} is a set $(\mathfrak{g}, [\,,\,], \delta)$, where $\mathfrak{g}$ is a Lie algebra\footnote{We will work with the field of complex numbers. We will also use for convenience anti-hermitian operators: $i\mathcal{O}\to \mathcal{O}$. Einstein's convention is used everywhere, with Greek indices running from 0 to 3 and Latin indices from 1 to 3.} defined by
\begin{equation}\label{cccc}
[X_i, X_j]={c_{ij}}^k X_k,\,\,\,\,\,\,\, X_i \in \mathfrak{g},\,\,\,\,\,\,\,{c_{ij}}^k\in \mathbb{C},
\end{equation}
where the structure constants ${c_{ij}}^k$ satisfy the Jacobi identity:
\begin{equation}\label{jacc}
{c_{ij}}^l{c_{lk}}^m+{c_{ki}}^l{c_{lj}}^m+{c_{jk}}^l{c_{li}}^m=0.
\end{equation}
$\delta:\mathfrak{g}\to \mathfrak{g}\otimes\mathfrak{g}$ is a skew-symmetric map, the \emph{cocommutator},  obeying a \emph{cocycle condition:}
\begin{equation}\label{cocond}
\delta([X_i, X_j])=[\delta(X_i), X_j\otimes 1 +1\otimes X_j]+[X_i\otimes1+1\otimes X_i,\delta(X_j) ] .
\end{equation}
Moreover the dual map $\delta^*:\mathfrak{g}^*\otimes\mathfrak{g}^*\to \mathfrak{g}^* $ is required to be a Lie bracket, making $\mathfrak{g}^*$ into a Lie algebra:
\begin{equation}
[\xi^i, \xi^j]={f^{ij}}_k \xi^k, \,\,\,\,\,\,\, \xi^i \in \mathfrak{g}^*,\,\,\,\,\,\,\,{f_{ij}}^k\in \mathbb{C},\,\,\,\,\,\,\, \langle \xi^i, X_j\rangle={\delta^i}_j,
\end{equation}
therefore we can expand the cocommutator on a basis
\begin{equation}\label{popo}
\delta(X_k)= {f^{ij}}_k X_i\wedge X_j .
\end{equation}
The structure constants $f$  obey a co-Jacobi identity (Jacobi identity for the dual algebra $\mathfrak{g}^*$):
\begin{equation}\label{cojacc}
{f^{jk}}_i{f^{lm}}_j+{f^{jl}}_i{f^{mk}}_j+{f^{jm}}_i{f^{kl}}_j=0,
\end{equation}
we can thus write the cocycle condition \eqref{cocond} in a more explicit manner
\begin{equation}\label{ccccccc}
{f^{ab}}_k{c_{ij}}^k={f^{ak}}_i{c_{kj}}^b+{f^{kb}}_i{c_{kj}}^a+{f^{ak}}_j{c_{ik}}^b+{f^{kb}}_j{c_{ik}}^a,
\end{equation}
this identity can be seen as a compatibility condition between  $\mathfrak{g}$ and  $\mathfrak{g}^*$ as  Lie bialgebra elements.

It so happens that if $\mathfrak{g}$ is semisimple then one can construct $\delta$ in an easier way. A Lie bialgebra is said to be a  coboundary if there exists an element 
\begin{equation}
r=r^{ij} X_i\wedge X_j ~~~ \in  ~~~ \mathfrak{g}\wedge\mathfrak{g},
\end{equation}
the ``$r$-matrix", such that
\begin{equation}\label{cobo}
\delta(X_i)=[X_i\otimes 1 +1 \otimes X_i, r], ~~~ X_i \in \mathfrak{g},
\end{equation}
which is always true when $\mathfrak{g}$ is semisimple, as a consequence of Whitehead's lemma~\cite{laurent2012poisson}. Furthermore in a coboundary Lie bialgebra $r$ is a solution of the \emph{Modified Classical Yang-Baxter equation} (mCYBE):
\begin{equation}\label{mCYBE}
[X_i\otimes1\otimes1+1\otimes X_i\otimes1+1\otimes1\otimes X_i, [[r, r]]\,]=0,
\end{equation}
where $[[r, r]]:= [r_{12}, r_{13}]+[r_{12}, r_{23}]+[r_{13}, r_{23}]\,$ is the \emph{Schouten bracket}. Here  $r_{12}=r^{ij} X_i\otimes 	X_j \otimes 1$ and the same convention is taken for $r_{13}$ and $r_{23}$.
If $[[r, r]]=0$ then $r$ is said to satisfy the \emph{Classical Yang-Baxter equation}  (CYBE).

As we said, Drinfel'd proved that a Hopf algebra $H$ can be specified (up to global issues) by its first order deformation, which is a Lie bialgebra~\cite{Drinfeld}.
Writing the Hopf algebra coproduct $\Delta: H \to H \otimes H$ as a series in powers of the generators,
one can in principle reconstruct higher order terms from the first order cocommutator $\delta$ by solving the co-associativity axiom 
order by order. 
Classifying Lie bialgebras is thus a key issue for listing all inequivalent quantum groups. This process is  easier if $\mathfrak{g}$ is semisimple, because then one has to look only for all  possible $r$ matrices. Lie bialgebras built from $\mathfrak{so}(d-1, 1)$ are always coboundaries since $\mathfrak{so}(d-1, 1)$ is semisimple, but in general this might not be true for any $\mathfrak{iso}(d-1, 1)$.
It can be nonetheless shown that all possible deformations of $\mathfrak{iso}(d-1, 1)$ are also coboundaries, for spacetime dimensions $d > 2$ \cite{zaz}.

\subsection{The meaning of coalgebraic structures: quantum spacetimes}
  
%

To have a better idea of the meaning of the additional  structures that  Lie bialgebras, consider a particular Lie-bialgebra deformation of the Poincar\'e  Lie algebra $\mathfrak{iso}(3, 1)$ in 3+1 dimensions. In the standard basis  $\{ P_0, P_i, K_i, J_i \}$ (which represent linear and angular momenta) the commutators take the form:
\begin{equation}\label{PoincareAlgebra}
\begin{aligned}
&[P_0, P_i]=0 \,, & &[P_0, K_i]=-P_i\,,& &[P_0, J_i]=0\,,&
\\
&[P_i, P_j]=0\,,& &[J_i, P_j]=\varepsilon_{ijk}P_k\,,& &[J_i, K_j]=\varepsilon_{ijk}K_k \,,&
\\
&[K_i, K_j]=-\varepsilon_{ijk}J_k\,,& &[J_i, J_j]=\varepsilon_{ijk}J_k\,& &[P_i, K_j]=\delta_{ij}P_0\,.&
\end{aligned}
\end{equation}
The so-called $\kappa$-deformation~\cite{lukkkk,mr,waves,Aschieri2017, kappakappa, aschierigeom} (which will feature prominently in the rest of the paper) is generated by the $r$-matrix: $
r = {\frac 1 \kappa} K_i\wedge P_i$. Using  \eqref{cobo} we obtain 
\begin{equation}\label{kappaPoincareBialgebra}
\begin{aligned}
&\delta(P_0)=0,& &\delta(P_i)= {\frac 1 \kappa} P_i\wedge P_0,&  &\delta(K_i)= {\frac 1 \kappa} (K_i\wedge P_0 +\varepsilon_{ijk}P_j\wedge J_k),&  &\delta(J_i)=0.
\end{aligned}
\end{equation}
What is the meaning of the above relations? To answer, consider the dual algebra $\mathfrak{iso}(3, 1)^*$, generated by the ten basis elements $\{a^\mu, \omega^\mu{}_\nu \}$. This algebra is, in the undeformed case, the \emph{algebra of functions over an infinitesimal neighbourhood around the identity of the Poincar\'e group}~\cite{flaaaaaaa}. The basis elements are to be understood as \emph{coordinate functions} on the group manifold (\emph{i.e.} $a^\mu$ are four functions which associate to a group element the corresponding translation vector in the standard representation).  The commutation relations~(\ref{PoincareAlgebra}) are dual to a set of rules 
$a^\mu \to a^\nu \wedge \omega^\mu{}_\nu$, $\omega^\mu{}_\nu \to\omega^\mu{}_\rho \wedge \omega^\rho{}_\nu$, which encode the way two infinitesimal Poincar\'e transformation combine (\emph{i.e.} they encode the group product).

In the deformed case, we can consider the dual of the $\kappa$-deformed Lie bialgebra, and the novelty is that the cocommutator map~(\ref{kappaPoincareBialgebra}) dualizes to a set of commutation relations for the coordinate functions:\footnote{We define $\xi^i = \omega^i{}_0$ and $\omega^i= {\frac 1 2} \varepsilon^{ijk} \omega_{jk}$.}
%
%
%
%
\begin{equation}\label{kappaPoincareDualBialgebra}
\begin{aligned}
&[x^0, x^i]=-{\frac 1 \kappa} x^i,& &[x^0, \xi^i]=- {\frac 1 \kappa}\xi^i,& &[x^0, \omega^i]=0, \\&
[x^i, x^j]=0,& &[x^i, \xi^j]=0,& &[x^i, \omega^j]={\frac 1 \kappa} \varepsilon_{ijk}\omega^k, \\&
[\xi^i, \xi^j]=0,& &[\xi^i, \omega^j]=0,& &[\omega^i, \omega^j]=0.
\end{aligned}
\end{equation}
The algebra of functions on the group manifold [which was commutative, being endowed with the pointwise product between functions, $(f \cdot g) (x)= f(x) g(x) = g(x) f(x) = (g \cdot f) (x)$] generalized to a nonabelian algebra. The group manifold does not admit anymore the interpretation of a topological manifold; it is instead, a sort of `quantum manifold': a \emph{noncommutative geometry}. In the case of the $\kappa$-deformation, the algebra~\eqref{kappaPoincareDualBialgebra} admits a subalgebra of translations:
\begin{equation}
[ x^0 , x^i ] = - \frac{1}{\kappa} x^i \,, \qquad [ x^i , x^j ] = 0  \,,
\end{equation}
which can be interpreted as the algebra of coordinates on a 4-dimensional noncommutative spacetime, known as \emph{$\kappa$-Minkowski}.\footnote{The fact that translation coordinates close a subalgebra is a property of the Lie bialgebra called \emph{coisotropy}. Together with other properties (which $\kappa$-Poincar\'e satisfies), it defines the notion of \emph{quantum homogenous space}~\cite{AngelFlavioUpcomingPaper}.}

Notice that the Lie bialgebra structures can at best deliver a \emph{Lie algebra} of noncommutative coordinates, \emph{i.e.} the commutators have to be linear. In full generality, one can expect commutation relations which are arbitrary functions of the coordinates, and indeed that is what happens with the full quantum group in $\kappa$-Poincar\'e, of which the Lie bialgebra is but the first order in a power expansion (see also~\cite{LukierskiQuadratic}, where a noncommutative spacetime with quadratic commutation relations is shown, which in our formalism, at the Lie bialgebra level, would appear commutative). Nevertheless, the lowest order of this expansion is expected to be the least suppressed by the physical constants that control the quantum deformation (see next Section), and in this paper we are concerned with leading-order effects.

\section{Lie-bialgebra Deformations of Relativistic Lie algebras}\label{bia}

In this Section we shall describe our approach for seeking and classifying Lie-bialgebra deformations. We call a \textit{deformation} of a Lie algebra with structure constants  ${c_{ij}}^k$ a cocommutator map satisfying \eqref{cojacc} and \eqref{ccccccc}. The goal of this Section is to investigate, through a systematic approach, what Lie bialgebra candidates can serve as generalizations of the algebras of homogeneous spacetimes symmetries. We shall find that the well-known $\kappa$-Poincar\'e group \cite{Lukierski1994} (with its related algebra of coordinates $\kappa$-Minkowski) is, under certain assumptions, the unique Hopf algebra extending the classic Poincar\'e Lie group to a noncommutative framework in $(3+1)$D. 

Consider the following Lie algebra 
$\mathfrak{g}_\Lambda$:
\begin{equation}\label{Poincare/dS/AdS-algebra} 
\begin{aligned}
&[M^{\mu\nu}, M^{\rho\sigma}] = \eta^{\nu\rho}M^{\mu\sigma}-\eta^{\mu\rho}M^{\nu\sigma}-\eta^{\sigma\mu}M^{\rho\nu}+\eta^{\sigma\nu}M^{\rho\mu} \,, 
\\
&[P^\mu,M^{\rho\sigma}] = \eta^{\mu\rho}P^\sigma-\eta^{\mu\sigma}P^\rho \,, \qquad [P^\mu, P^\nu] = - \Lambda ~ M^{\mu\nu} \,,
\end{aligned}
\end{equation}
splitting space ($i,j,k,\dots=1,2,3$) and time indices, and introducing the boost and rotation generators:
\begin{equation}
M^{0i}=K_i \,, \qquad J_i=-\frac{1}{2}\varepsilon_{ijk}M^{jk} \,,
\qquad
M^{ij} = - \varepsilon^{ijk} J_k \,,
\end{equation}
which mean that $J_1= - M^{23}$, $J_2 = - M^{31}$ and $J_3 = - M^{12}$, with the following convention for the Minkowski metric: $\eta^{00} = -1$ and $\eta^{ij} = \delta^{ij}$, we get the following commutation relations:
\begin{equation}\label{(A)dS_algebra_3+1_space_indicesson}
\begin{aligned}
&[P_0, P_i]=-\Lambda  K_i \,, & &[P_0, K_i]=-P_i\,,& &[P_0, J_i]=0\,,&
\\
&[P_i, P_j]=\Lambda  \varepsilon_{ijk} J_k\,,& &[J_i, P_j]=\varepsilon_{ijk}P_k\,,& &[J_i, K_j]=\varepsilon_{ijk}K_k \,,&
\\
&[K_i, K_j]=-\varepsilon_{ijk}J_k\,,& &[J_i, J_j]=\varepsilon_{ijk}J_k\,& &[P_i, K_j]=\delta_{ij}P_0\,.&
\end{aligned}
\end{equation}


The algebra above corresponds to  the Poincar\'e algebra $\mathfrak{iso}(3,1)$ when $\Lambda=0$, to the de Sitter algebra $\mathfrak{so}(4,1)$ when $\Lambda>0$ and to the Anti-de Sitter algebra $\mathfrak{so}(3,2)$ when $\Lambda<0$.
In order to deform \eqref{Poincare/dS/AdS-algebra} we introduce the cocommutator map $\delta :  \mathfrak{g}_\Lambda \to \mathfrak{g}_\Lambda \wedge \mathfrak{g}_\Lambda$, whose most general form we write as
\begin{equation}\label{General_cocommutator} 
\begin{split}
\delta(P_\mu) &= \mathcal{A}_\mu^{\:\: \rho\sigma}P_\rho\wedge P_\sigma+\mathcal{B}_\mu^{\:\:\rho\sigma\gamma}P_\rho\wedge M_{\sigma\gamma}+\mathcal{C}_\mu^{\:\:\rho\sigma\gamma\delta} M_{\rho\sigma}\wedge M_{\gamma\delta}, \\
\delta(M_{\mu\nu}) &= \mathcal{D}_{\mu\nu}^{\:\:\:\: \rho\sigma}P_\rho\wedge P_\sigma+\mathcal{E}_{\mu\nu}^{\:\:\:\:\rho\sigma\gamma}P_\rho\wedge M_{\sigma\gamma}+\mathcal{F}_{\mu\nu}^{\:\:\:\:\rho\sigma\gamma\delta} M_{\rho\sigma}\wedge M_{\gamma\delta},
\end{split}
\end{equation}
and impose algebraic conditions on $\delta$  that define a Lie bialgebra: Jacobi identity on $ \mathfrak{g}^*_\Lambda$ structure constants (co-Jacobi identity) and cocycle condition \eqref{ccccccc}. The coefficients $\{\mathcal{A, B, C, D, E, F}\}$ will be functions of the physical constants $L_p$ and $\Lambda$, chosen on the basis of certain symmetry requirements.
  
\subsection{Dimensional Analysis}\label{DimAn}

If we work in units such that $c=\hbar=1$,  the translation generators $P_\mu$ of \eqref{Poincare/dS/AdS-algebra} have the dimensions of an energy, the Lorentz generators $M_{\mu\nu}$ are dimensionless and the cosmological constant $\Lambda$ has dimensions of energy squared.

We can make the $P_\mu$ generators dimensionless by dividing them by the square root of the norm of the cosmological constant:
\begin{equation}
P_\mu = \sqrt{|\Lambda|} \, Q_\mu\,,
\end{equation}
then, if we introduce
\begin{equation}
\lambda = \text{sign}(\Lambda) \,,
\end{equation}
(with the convention that $\Lambda =0$ $\Rightarrow$ $\lambda =0$) we can make the whole algebra  \eqref{Poincare/dS/AdS-algebra} dimensionless:
\begin{equation}\label{Dimensionlessbialgebra} 
\begin{split}
[M^{\mu\nu}, M^{\rho\sigma}] &= \eta^{\nu\rho}M^{\mu\sigma}-\eta^{\mu\rho}M^{\nu\sigma}-\eta^{\sigma\mu}M^{\rho\nu}+\eta^{\sigma\nu}M^{\rho\mu},\\
[Q^\mu,M^{\rho\sigma}] &= \eta^{\mu\rho}Q^\sigma-\eta^{\mu\sigma}Q^\rho, \qquad [Q^\mu, Q^\nu] =-\lambda M^{\mu\nu}.
\end{split}
\end{equation}

If we now  write the most general cocommutator with the dimensionless variables $Q_\mu$, $M_{\mu\nu}$, this will be of the form:
\begin{equation}
\begin{split}
\delta(Q_\mu) &=  a_\mu^{\:\: \rho\sigma}Q_\rho\wedge Q_\sigma+b_\mu^{\:\:\rho\sigma\gamma}Q_\rho\wedge M_{\sigma\gamma}+c_\mu^{\:\:\rho\sigma\gamma\delta} M_{\rho\sigma}\wedge M_{\gamma\delta} , \\
\delta(M_{\mu\nu}) &=  d_{\mu\nu}^{\:\:\:\: \rho\sigma}Q_\rho\wedge Q_\sigma+ e_{\mu\nu}^{\:\:\:\:\rho\sigma\gamma}Q_\rho\wedge M_{\sigma\gamma}+f_{\mu\nu}^{\:\:\:\:\rho\sigma\gamma\delta} M_{\rho\sigma}\wedge M_{\gamma\delta},
\end{split}
\end{equation}
where the  coefficients ${a_\mu}^{\rho\sigma},...\,, {f_{\mu\nu}}^{\rho\sigma\gamma\delta}$  are dimensionless. We assume these coefficients to be  analytic functions of the only two physical scales in the model: the Planck length $L_p$ and the cosmological radius $\frac 1 {\sqrt{|\Lambda|}}$. There is only one way to make a dimensionless constant out of those two, and it is to take the combination 
\begin{equation}\label{qq}
\q =  L_p\sqrt{|\Lambda|} ,
\end{equation}
Then the coefficients will have to be analytic functions of the dimensionless parameter $\q$ (the ratio between the Planck length and the cosmological radius).
We require, as a physical input, that $\delta \xrightarrow[L_p \to 0]{} 0$, \emph{i.e.} that in the limit in which the Planck length vanishes, the coefficients ${a_\mu}^{\rho\sigma}(\q),...\,, {f_{\mu\nu}}^{\rho\sigma\gamma\delta}(\q)$ vanish too. Then the assumption of analyticity allows us to expand the coefficients in Taylor series around zero:
\begin{equation}
\begin{aligned}
{a_\mu}^{\rho\sigma}(\q) &=  \q \, a_\mu^{(1)\rho\sigma} + {\frac 1 2} \q^2 \, a_\mu^{(2) \rho\sigma} + \mathcal O(\q^3) \,,
\\
&\vdots
\\
{f_{\mu\nu}}^{\rho\sigma\gamma\delta}(\q) &=  \q \, f_\mu^{(1)\rho\sigma} +  {\frac 1 2} \q^2 f_\mu^{(2)\rho\sigma} + \mathcal O(\q^3) \,,
\end{aligned}
\end{equation}
where $a_\mu^{(i)\rho\sigma}$, \dots , $f_{\mu\nu}^{(i)\rho\sigma\gamma\delta}$ are numerical. Then Eq.~\eqref{General_cocommutator} reads, at first order in $\q$:
\begin{equation}
\begin{split}
\delta(Q_\mu) &=  \q  \left( a_\mu^{(1) \rho\sigma}Q_\rho\wedge Q_\sigma+b_\mu^{(1)\rho\sigma\gamma}Q_\rho\wedge M_{\sigma\gamma}+c_\mu^{(1)\rho\sigma\gamma\delta} M_{\rho\sigma}\wedge M_{\gamma\delta} \right) + \mathcal O(\q^2) , \\
\delta(M_{\mu\nu}) &=  \q  \left( d_{\mu\nu}^{(1)\rho\sigma}Q_\rho\wedge Q_\sigma+ e_{\mu\nu}^{(1)\rho\sigma\gamma}Q_\rho\wedge M_{\sigma\gamma}+f_{\mu\nu}^{(1)\rho\sigma\gamma\delta} M_{\rho\sigma}\wedge M_{\gamma\delta} \right) + \mathcal O(\q^2) ,
\end{split}
\end{equation}
and reintroducing the dimensionful translation generators by $Q_\mu = \frac{P_\mu}{\sqrt{|\Lambda|}} $: 
\begin{equation}\label{dimm}
\begin{split}
\delta(P_\mu) &=   L_p  \left(  a_\mu^{(1) \rho\sigma}P_\rho\wedge P_\sigma+ \sqrt{|\Lambda|}  \, b_\mu^{(1)\rho\sigma\gamma}P_\rho\wedge M_{\sigma\gamma}+|\Lambda| \, c_\mu^{(1)\rho\sigma\gamma\delta} M_{\rho\sigma}\wedge M_{\gamma\delta} \right) + \mathcal O(\q^2) , \\
\delta(M_{\mu\nu}) &= L_p  \left( \frac{d_{\mu\nu}^{(1)\rho\sigma}}{ \sqrt{|\Lambda|} } P_\rho\wedge P_\sigma+  e_{\mu\nu}^{(1)\rho\sigma\gamma}P_\rho\wedge M_{\sigma\gamma}+  \sqrt{|\Lambda|}  \, f_{\mu\nu}^{(1)\rho\sigma\gamma\delta} M_{\rho\sigma}\wedge M_{\gamma\delta} \right) + \mathcal O(\q^2) .
\end{split}
\end{equation}
this shows the relationship between the ${\mathcal{A}_\mu}^{\rho\sigma}$, \dots , ${{\mathcal{F}}_{\mu\nu}}^{\rho\sigma\gamma\delta}$ coefficients and the Taylor expansion:
\begin{equation}\label{taylor}
\begin{split}
&\mathcal{A}_\mu^{\:\: \rho\sigma} = L_p  \,   a_\mu^{(1)\rho\sigma}  + L_p^2 \, \sqrt{|\Lambda|}  \,   a_\mu^{(2)\rho\sigma} + \mathcal O(L_p^3)
\\
&\mathcal{B}_\mu^{\:\:\rho\sigma\gamma} =  L_p \, \sqrt{|\Lambda|}  \, b_\mu^{(1)\rho\sigma\gamma}  +  L_p^2 \, |\Lambda| \, b_\mu^{(2)\rho\sigma\gamma}  + \mathcal O(L_p^3) \,,
\\
&\mathcal{C}_\mu^{\:\:\rho\sigma\gamma\delta}  =  L_p \, |\Lambda| \, c_\mu^{(1)\rho\sigma\gamma\delta} +  L_p^2 \, |\Lambda|^{\frac 3 2} c_\mu^{(2)\rho\sigma\gamma\delta}+ \mathcal O(L_p^3) \,,
\\
&\mathcal{D}_{\mu\nu}^{\:\:\:\: \rho\sigma} = \frac{L_p }{\sqrt{|\Lambda|}} \, d_{\mu\nu}^{(1)\rho\sigma} + L_p^2 \, d_{\mu\nu}^{(2)\rho\sigma}+ \mathcal O(L_p^3) \,,
\\
&\mathcal{E}_{\mu\nu}^{\:\:\:\:\rho\sigma\gamma} = L_p \, e_{\mu\nu}^{(1)\rho\sigma\gamma} + L_p^2 \,  \sqrt{|\Lambda|} \, e_{\mu\nu}^{(2)\rho\sigma\gamma}+ \mathcal O(L_p^3) \,,
\\
&\mathcal{F}_{\mu\nu}^{\:\:\:\:\rho\sigma\gamma\delta} =  L_p \, \sqrt{|\Lambda|}  \,  f_{\mu\nu}^{(1)\rho\sigma\gamma\delta}  + L_p^2 \, |\Lambda| \, f_{\mu\nu}^{(2)\rho\sigma\gamma\delta} + \mathcal O(L_p^3) \,.
\end{split}
\end{equation}
A few observations can be deduced from the expression above. First of all, the coefficient $\mathcal{D}_{\mu\nu}^{\:\:\:\: \rho\sigma}$ \emph{does not admit a regular flat limit} $\Lambda \to 0$, unless we set $d_{\mu\nu}^{(1)\rho\sigma} =0$. Then this coefficient is either second order in the Planck length or it is excluded by the reasonable requirement of admitting a flat limit. Moreover, the coefficients $\mathcal{B}_\mu^{\:\:\rho\sigma\gamma}$, $\mathcal{C}_\mu^{\:\:\rho\sigma\gamma\delta} $ and  $\mathcal{F}_{\mu\nu}^{\:\:\:\:\rho\sigma\gamma\delta}$ all start with a nonzero power of $ \sqrt{|\Lambda|}$. Therefore (at first order in $L_p$) they represent \emph{infrared corrections} to the deformation given by the coefficients $\mathcal{A}_\mu^{\:\: \rho\sigma}$ and $\mathcal{E}_{\mu\nu}^{\:\:\:\:\rho\sigma\gamma}$, and they all vanish in the flat limit $\Lambda \to 0$. For these reasons, if we are interested in the Lie-bialgebra deformations of the Poincar\'e algebra at first order in $L_p$, we can focus on the coefficients $\mathcal{A}_\mu^{\:\:\rho\sigma}$ and $\mathcal{E}_{\mu\nu}^{\:\:\:\:  \rho\sigma\gamma} $. However we do not want to throw all the other terms from the beginning; for visual convenience we will color the different terms differently. From now on, we indicate in blue the coefficients $\mathcal{A}_\mu^{\:\:\rho\sigma}$, $\mathcal{E}_{\mu\nu}^{\:\:\:\:  \rho\sigma\gamma} $ and their corresponding terms in $\mathfrak{g}_\Lambda  \wedge \mathfrak{g}_\Lambda $. The coefficients $\mathcal{B}_\mu^{\:\:\rho\sigma\gamma}$ and $\mathcal{F}_{\mu\nu}^{\:\:\:\:\rho\sigma\gamma\delta}$ will be corrections of order $\sqrt{|\Lambda|}$, and $\mathcal{C}_\mu^{\:\:\rho\sigma\gamma\delta}$ of order $|\Lambda|$, to the previous terms, and we will color them in red. We will leave the terms $\mathcal{D}_{\mu\nu}^{\:\:\:\:\rho\sigma}$ in black. The general cocommutator will then look like:
\begin{equation}\label{General_cocommutator2} 
\begin{split}
\delta(P_\mu) &= \blue{\mathcal{A}_\mu^{\:\: \rho\sigma}P_\rho\wedge P_\sigma} + \red{\mathcal{B}_\mu^{\:\:\rho\sigma\gamma}P_\rho\wedge M_{\sigma\gamma}} + \red{\mathcal{C}_\mu^{\:\:\rho\sigma\gamma\delta} M_{\rho\sigma}\wedge M_{\gamma\delta}}, \\
\delta(M_{\mu\nu}) &= \mathcal{D}_{\mu\nu}^{\:\:\:\: \rho\sigma}P_\rho\wedge P_\sigma+\blue{\mathcal{E}_{\mu\nu}^{\:\:\:\:\rho\sigma\gamma}P_\rho\wedge M_{\sigma\gamma}}+\red{\mathcal{F}_{\mu\nu}^{\:\:\:\:\rho\sigma\gamma\delta} M_{\rho\sigma}\wedge M_{\gamma\delta}}.
\end{split}
\end{equation}
\begin{table}[h]\centering
\begin{tabular}{lll}
\hline
color & lowest order & next-to-lowest order
\\$\,\,$
\blue{$\blacksquare$} &  $L_p$ & $L_p^2 \sqrt{|\Lambda|}$
\\
$\,\,$
\red{$\blacksquare$} & $L_p \sqrt{|\Lambda|}$ & $L_p^2  |\Lambda|$
\\
$\,\,$
$\blacksquare$ & ${L_p}/{\sqrt{|\Lambda|}}$ & $L_p^2$
\\
\hline
\end{tabular}
\caption{Color legend for the cocommutator coefficients.}
\end{table}

\subsubsection{Comparison With a Different Approach}

In the previous Section we were able to draw conclusions on the ``flat-limit" properties of a Lie bialgebra by dimensional analysis.
A somewhat similar result can be obtained by group contraction techniques applied to Lie bialgebras~\cite{balles95}. If $\phi_\varepsilon$ is a one-parameter family of Lie algebra automorphisms , one can define a contracted cocommutator $\delta'$ by:
\begin{equation}\label{contr}
\delta ' :=\lim_{\varepsilon\to 0} \varepsilon^n ({\phi^{-1}_\varepsilon}\otimes {\phi^{-1}_\varepsilon})\circ\delta\circ \phi_\varepsilon\,,
\end{equation}
if there is an $n$ such that this limit exists. Furthermore there is a minimal value $n_0$ of $n$ such that for $n\geq n_0$ the limit \eqref{contr} exists, and if $n>n_0$ it is zero. The automorphism used in \cite{balles95} are of the form $\phi_{\varepsilon} (J)= \varepsilon J$, where $\varepsilon$ is related to the algebra structure constants $f$: $\varepsilon=\sqrt{f}$. The connection between the two techniques can be understood by identifying $q=L_p\sqrt{|\Lambda|} $ of eq. \eqref{qq} with $\varepsilon$. The minimal value $n_0$ would then be the smallest power of $q$ such that the flat limit $\Lambda\to 0$ in \eqref{dimm} exists. For instance, looking at \eqref{taylor}, $n_0=2$ if we do not require $ d_{\mu\nu}^{(1)\rho\sigma}=0$.

We now turn to introducing the simplifying ans\"atze that will allow us to reduce the number of solutions  of the Lie-bialgebra conditions~\eqref{cojacc} and~\eqref{ccccccc} to a manageable amount. We begin with the strongest assumptions (manifest spacetime covariance), and proceed towards weaker assumptions: manifest spatial isotropy, $\mP$ and $\mT$ covariance and, finally, we will discuss what we know of the Lie-bialgebra deformations of spacetime symmetry algebras in full generality.

\subsection{Manifest spacetime covariance, spatial isotropy, $\mP$ and $\mT$ involutions}

The strongest simplifying assumption we can make is to assume that the cocommutator is covariant \emph{in form} under Lorentz transformations. This means that in the expression~\eqref{General_cocommutator} all the ${\mathcal{A}_\mu}^{\rho\sigma}$, \dots , ${\mathcal{F}_{\mu\nu}}^{\rho\sigma\gamma\delta}$ coefficients are combinations of invariant tensors with all indices saturated\footnote{Except the ``external" ones, which saturate with algebra generators.}. Inspecting the algebra $\mathfrak{g}_\Lambda$ in Eq.~\eqref{Poincare/dS/AdS-algebra}, we see that, independently of the dimension and of the sign of $\Lambda$, the Lorentz transformations act on each other and on the translation generators as they do in flat space (\emph{i.e.} as infinitesimal hyperbolic rotations). Therefore, if we want to form Lorentz-covariant combinations of the generators $P_\mu$ and $M_{\mu\nu}$, we can only use the two invariant tensors on Minkowski space: the flat metric $\eta^{\mu\nu} = \text{diag}\{ -1 , 1,\dots,1 \}$, and the totally antisymmetric Levi-Civita tensor $\varepsilon^{\mu \dots \sigma}$, such that $\epsilon^{01\dots(d-1)}=+1$. We can raise the indices with the metric, and put the general cocommutator in the following form:
\begin{equation}\label{General_cocommutator3} 
\begin{split}
\delta(P^\mu) &= \blue{\mathcal{A}^{\mu\rho\sigma}P_\rho\wedge P_\sigma} + \red{\mathcal{B}^{\mu\rho\sigma\gamma}P_\rho\wedge M_{\sigma\gamma}} + \red{\mathcal{C}^{\mu\rho\sigma\gamma\delta} M_{\rho\sigma}\wedge M_{\gamma\delta}}, \\
\delta(M^{\mu\nu}) &= \mathcal{D}^{\mu\nu\rho\sigma}P_\rho\wedge P_\sigma+\blue{\mathcal{E}^{\mu\nu\rho\sigma\gamma}P_\rho\wedge M_{\sigma\gamma}}+\red{\mathcal{F}^{\mu\nu\rho\sigma\gamma\delta} M_{\rho\sigma}\wedge M_{\gamma\delta}},
\end{split}
\end{equation}
then each coefficient has only contravariant indices, and inherits certain symmetry properties from the antisymmetry of the wedge product and of $M_{\mu\nu}$: $\blue{\mathcal{A}^{\mu\rho\sigma}}$ and $\mathcal{D}^{\mu\nu\rho\sigma}$ are antisymmetric in the indices $\rho$, $\sigma$;  $\red{\mathcal{B}^{\mu\rho\sigma\gamma}}$ and $\blue{\mathcal{E}^{\mu\nu\rho\sigma\gamma}}$ are antisymmetric in $\sigma$, $\gamma$, while $\red{\mathcal{C}^{\mu\rho\sigma\gamma\delta}}$ and $\red{\mathcal{F}^{\mu\nu\rho\sigma\gamma\delta}}$ are completely antisymmetric in $\rho$, $\sigma$, $\gamma$ and $\delta$.

For each choice of dimensions we need to find all the independent invariant tensors with 3, 4, 5 and 6 indices, and write each coefficient $\mathcal{A}^{\mu\rho\sigma}$, \dots , $\mathcal{F}^{\mu\nu\rho\sigma\gamma\delta}$ as a linear combination of all the invariant tensors with the appropriate number of indices.

The following ansatz we can make, in order of strength, is a covariance in form under spatial rotations. Differentiating the time index $\mu = 0$ from the spatial ones $\mu = i = 1, \dots, d$, we introduce the generators of pure boosts, $K_i = M^{0i}$, and pure rotations, $J_i = - \frac 1 2 \epsilon_{ijk} M^{jk}$. Then manifest spatial isotropy means that all the spatial indices in the Lie bialgebra are saturated correctly and match on both sides. So for example the cocommutator of $P_0$ will have to be a scalar (index-wise), and therefore will contain terms like $P_i\wedge P^i$, $P_i\wedge K^i$ \emph{etc.} depending on the spatial dimension $d$, we have at our disposal, on top of the spatial Euclidean metric $\delta_{ij}$, also a rank-$d$ tensor, the Levi-Civita tensor $\epsilon^{i_i \dots i_d}$, to saturate the indices on the right-hand side. Then the kind of terms we may form are different in 2 and 3 spatial dimensions. In what follows we will write explicitly the spatially-isotropic ansatz in 2+1 and 3+1 dimensions.

Finally, one might be interested in studying the discrete symmetries of a noncommutative QFT. To this end a natural request is that the discrete transformations are involutions which are Lie-bialgebra automorphisms (for a detailed discussion on involutions and quantum algebras see for instance \cite{spacelike}). In a fully nonlinear Hopf-algebraic setting, one might expect these discrete transformations to get nonlinear (\emph{e.g.} energy-dependent-) corrections, as was suggested first in~\cite{waves} and further developed in~\cite{cpt}. In particular, these works propose that the antipode map of the quantum group should be involved in the definition of discrete transformations.\footnote{This is not unproblematic in general: in the case considered by~\cite{cpt} (\emph{i.e.} $\kappa$-Poincar\'e), the antipode is not an involution for the full group, but only for the translation subgroup, and this seems to lead to a non-Lorentz-invariant notion of $\mT$ transformations.}
However we are working at the Lie bialgebra level, and we are not sensitive to nonlinear structures. The antipode is a nonlinear map which can be linearized in a neighbourhood of the identity, and its linear order always reduces to the trivial involution that flips all the signs of the algebra generators (exactly for the same reason that the Lie group inverse reduces, at the linear level, to a sign change of all Lie algebra generators). So, for instance, the parity-transformed spatial momentum used iny~\cite{cpt} can be written as:
\begin{equation}
\mP_\kappa (P_i)=S(P_i)=-P_i+\mathcal{O}\left(\frac 1 \kappa \right),
\end{equation}
which is an expression that, in the low-energy/momentum limit\footnote{Or, conversely, in the limit in which the deformation parameters go to zero.} reduces to the familiar parity operator of the undeformed Poincar\'e algebra. In order for a parity/time-reversal operator to exist at the level of the quantum group, it is necessary for the corresponding zeroth-order operator to leave the Lie bialgebra invariant. This necessary condition is what we focus on in the present paper.\footnote{One could imagine a Hopf-algebra deformation of Poincar\'e/(A)dS in which the discrete symmetries reduce to the ordinary one in the limit of vanishing deformation parameter, \emph{e.g.} $\kappa \to \infty$, but they do not do so in their linearized form. For example, dimensional analysis allows us to write expressions like
$\mP (K_i)= - K_i+ \frac 1 \kappa P_i$. This expression reduces to $\mP (K_i)= - K_i$ in the $\kappa \to \infty$ limit, but the deformation term is not nonlinear, and survives the linearization procedure that defines the Lie bialgebra. We will postpone the investigation of this peculiar case to future works.} The transformations we consider are spatial inversions (parity, or $\mP$) and time-inversions ($\mT$).
%
We define  $\mP$ as
\begin{equation}\label{d1}
\mP(P_0)=P_0, \qquad \mP(J_i)=J_i, \qquad \mP(P_i)=-P_i, \qquad \mP(K_i)=-K_i,
\end{equation}
in what follows we will highlight $\mP$-covariant terms in a black box.
We will also look for solutions (highlighted in a red box) respecting a $\mT$-inversion, defined as:
\begin{equation}\label{d2}
\mT(P_0)=-P_0, \qquad \mT(J_i)=J_i, \qquad \mT(P_i)=P_i, \qquad \mT(K_i)=-K_i.
\end{equation}
 As we will see these requirements impose nontrivial restrictions on Lie bialgebra solutions of all dimensions.

\section{Lie Bialgebra solutions}\label{sol}

To find Lie-bialgebra deformations we have to solve Eqs.~\eqref{ccccccc} and~\eqref{cojacc} for the structure constants  $ {f^{ij}}_k$, with ${c_{ij}}^k$ fixed by~\eqref{Poincare/dS/AdS-algebra}. We first solve the cocycle condition~\eqref{ccccccc} which is linear in $ {f^{ij}}_k$ and therefore possesses always a unique solution. We then plug the solution  into the co-Jacobi relations~\eqref{cojacc} which are quadratic in $ {f^{ij}}_k$. In this way we take full advantage of the exact solvability of Eqs.~\eqref{ccccccc} to reduce the number of independent variables to the minimum. 

This strategy is, in the most general 3D and 4D cases, not feasible, because the nonlinearity of Eqs.~\eqref{cojacc} leads to a profusion of solutions which, although they may be tackled with a computer algebra program, are way too many to make sense of.

In order to overcome this issue we add  further constraint by specifying, through the physical requirements listed above, a particular ans\"atze for the variables ${f^{ij}}_k$.

\subsection{Two spacetime Dimensions}

\subsubsection{Manifest spacetime covariance}

In 1+1 dimensions both the Levi-Civita symbol and the metric have two indices. Therefore we can only form invariant tensors of even rank. There are only five independent terms:
\begin{equation}
\begin{split}
&\blue{\mathcal{A}^{\mu\rho\sigma}} = 0 \,, 
\\
&\red{\mathcal{B}^{\mu\rho\sigma\gamma}} = \red{c_1 \, \eta^{\rho\gamma} \eta^{\mu\sigma}} + \red{ c_2 \,\epsilon^{\rho\gamma}\eta^{\mu\sigma} }   \,,
\\
&\red{\mathcal{C}^{\mu\rho\sigma\gamma\delta}}  = 0   \,,
\\
&\mathcal{D}^{\mu\nu\rho\sigma} = c_5 \, \eta^{\mu\rho} \eta^{\nu\sigma}   \,,
\\
&\blue{\mathcal{E}^{\mu\nu\rho\sigma\gamma}} = 0  \,,
\\
&\red{\mathcal{F}^{\mu\nu \rho\sigma\gamma\delta}} = \red{ c_3 \, \eta^{\mu\rho} \eta^{\nu\delta} \eta^{\sigma\gamma}} + \red{ c_4 \, \eta^{\mu\rho} \eta^{\nu\delta} \epsilon^{\sigma\gamma} } \,.
\end{split}
\end{equation}
Imposing Jacobi identities and cocycle conditions gives $\red{c_1} =  \red{c_3} =  \red{c_4} =0$, $\lambda \, c_5 = 0$ and $\red{c_2} c_5 = 0$. There are then two nontrivial solutions (which reduce to one if the cosmological constant is nonzero:
\begin{equation}\label{1+1_spacetime_iso}
\begin{split}
\delta(P_\mu) &=  \red{ c_2 \,\epsilon^{\rho\sigma}  P_\rho \wedge M_{\mu\sigma}}  , 
\\
\delta(M_{\mu\nu}) &= 0 ,
\end{split}
\qquad
\begin{split}
\delta(P_\mu) &= 0,
\\
\delta(M_{\mu\nu}) &= \colorboxed{red}{\boxed{(1 - |\lambda| ) c_5 \,P_\mu \wedge P_\nu }} ,
\end{split}
\end{equation}
recall that we highlighted the $\mT$-invariant terms with a red box, and the $\mP$-invariant ones with a black box, so the $c_5$ term is invariant under both $\mT$ and $\mP$, while $\red{c_2}$ is invariant under neither.
The solutions in each case depend on only one free parameter (so they are essentially unique).

\subsubsection{Coboundary case}
All the algebras we consider are coboundaries, but only in spacetime dimensions greater than 2. Hence, imposing the coboundary condition is a restrictive ansatz only in the $(1+1)$D case.

The most general $r$-matrix is:
\begin{equation}
r = b_1 \, K_1 \wedge P_0 + b_2 \, K_1 \wedge P_1 + b_3 \, P_0 \wedge P_1 \,,
\end{equation}
where $K_1 = M_{01}$ is the boost generator. An $r$-matrix satisfies automatically the cocycle conditions. The only nontrivial conditions are the  co-Jacobi rules, which however in this case  only impose that if $\lambda=0$ then $b_3 = 0$. The resulting cocommutator is
\begin{equation}\label{1+1_coboundary}
\begin{split}
\delta(P_0) &= \colorboxed{red}{ \blue{b_1 \, P_0 \wedge P_1} }+\Lambda \, \red{b_3 \, K_1 \wedge P_0}  , \\
\delta(P_1) &= \boxed{- \blue{b_2 \, P_0 \wedge P_1}} +{\Lambda \,\red{b_3 \, K_1 \wedge P_1}} ,\\
\delta(K_1) &= \colorboxed{red}{ \blue{b_1 \, K_1 \wedge P_1}} +\boxed{ \blue{b_2 \, K_1 \wedge P_0}} .
\end{split}
\end{equation}

\subsubsection{General case}

Making a completely general ansatz:
\begin{equation*}
\begin{split}
\delta(P_0) &=  \red{c_1 \, K_1\wedge P_0} + \red{c_2 \, K_1\wedge P_1} + \blue{c_3 \,  P_0\wedge P_1}, 
\\
\delta(P_1) &=  \red{c_4 \, K_1\wedge P_0}+ \red{c_5 \,  K_1\wedge P_1} + \blue{c_6 \,  P_0\wedge P_1}.
\\
\delta(K_1) &=  \blue{c_7 \, K_1\wedge P_0} + \blue{c_8 \, K_1\wedge P_1} + c_9 \, P_0\wedge P_1, 
\end{split}
\end{equation*}    
and imposing cocycle and co-Jacobi conditions, we obtain the conditions $\red{c_2}=\blue{c_3}=\red{c_4}=\red{c_5}=\blue{c_6}=0$ and $\lambda c_9 =0$:
\begin{equation}
\begin{split}
\delta(P_0) &=  \red{c_1 \, K_1\wedge P_0} + \colorboxed{red}{\blue{c_8 \,  P_0\wedge P_1}}, 
\\
\delta(P_1) &=  \red{c_1 \,  K_1\wedge P_1} -  \boxed{ \blue{c_7 \,  P_0\wedge P_1}}.
\\
\delta(K_1) &=   \boxed{ \blue{c_7 \, K_1\wedge P_0}} +\colorboxed{red}{  \blue{c_8 \, K_1\wedge P_1} +  \boxed{ (1-|\lambda|)  c_9 \, P_0\wedge P_1}}, 
\end{split}
\end{equation}    
the above cocommutator reduces to a coboundary in the case $c_9=0$. To connect the $r$-matrix parameters written above with the parameters used here we can write:
\begin{equation}
 \red{c_1} = \Lambda \, \red{b_3}, \qquad \blue{c_7} = \blue{b_2}, \qquad \blue{c_8} = \blue{b_1}, \qquad  c_9 = 0 \,,
\end{equation}
this solution reduces to  the manifestly spacetime covariant case when $c_1=c_2, \,\,c_9=c_5$ and others $c_i=0.$

\subsection{Three spacetime Dimensions}

\subsubsection{Manifest spacetime covariance}

Since in 3 dimensions the basic invariant tensors are of rank 3 and 2, one can form invariant combinations of all ranks. A complete basis of tensor invariant expressed in terms of $\varepsilon_{\mu\nu\rho}$ and $\eta_{\mu\nu}$ can be found in the literature \cite{kear}. For instance the first two coefficients can be written as 
\begin{equation*}\label{eq:2} 
\begin{split}
&\blue{\mathcal{A}^{\mu\rho\sigma}} =  \blue{f_1 \, \varepsilon^{\mu\nu\rho} } ,
\\
&\red{\mathcal{B}^{\mu\rho\sigma\gamma}} = \red{f_2 \, \eta^{\mu\nu}\eta^{\rho\sigma}} +  \red{f_3 \, \eta^{\mu\rho}\eta^{\nu\sigma}} +  \red{f_4 \, \eta^{\mu\sigma}\eta^{\rho\nu}},
\end{split}
\end{equation*}
and analogous forms hold for the remaining terms. 

Imposing the cocycle and co-Jacobi conditions, it turns out that the most general solution for the cocommutators is unique and reads
\begin{equation}\label{2+1-spacetime_covariant} 
\begin{split}
\delta(P_\mu) &= \boxed{2 \, \blue{f_1\, \varepsilon_\mu^{\:\: \rho\sigma}P_\rho\wedge P_\sigma} +  2 \Lambda \red{\,  f_1 \, \varepsilon^{\rho\sigma\gamma}M_{\mu\rho} \wedge M_{\sigma\gamma}}}, \\
\delta(M_{\mu\nu}) &= 0.
\end{split}
\end{equation}
The result is $\mP$-invariant, but not $\mT$-invariant.

\subsubsection{Manifest spatial isotropy}\label{SEC_2+1_spatially-isotropic}

The most general manifestly spatially isotropic cocommutator expression in the 6-dimensional basis $\{P_0, J_3, P_i, K_i  \}$ (where $K_i = M_{0i}$, $J_3 = M_{12}$) is 
\begin{equation}\label{2+1-spatially_isotropic} 
\begin{split}
\delta(P_0) =&  \red{f_1 \, P_0 \wedge J_3} + \red{f_2 \, P_i \wedge K^i} + ( \red{f_3 \, P_i \wedge K_j } +  \blue{f_4  \,  P_i \wedge P_j} + \red{f_5 \,  K_i \wedge K_j} )\varepsilon_{ij}, \\
\delta(J_3) =& \blue{f_6 \, P_0 \wedge J_3} + \blue{f_7 \, P_i \wedge K^i} + ( \blue{f_8 \, P_i \wedge K_j } +  f_9  \,  P_i \wedge P_j + \red{f_{10} \,  K_i \wedge K_j} )\varepsilon_{ij}, \\
\delta(P_i) =&  \blue{f_{11} \, P_0 \wedge P_i} + \red{f_{12} \, P_0 \wedge K_i} +\red{f_{13} \, J_3 \wedge K_i} + \red{f_{14} \, J_3 \wedge P_i} +
\\
& \left( \blue{f_{15} \, P_0 \wedge P_j} + \red{f_{16} \, P_0 \wedge K_j} + \red{f_{17} \, J_3 \wedge K_j} + \red{f_{18} \, J_3 \wedge P_j} \right) \varepsilon_{ij} ,\\
\delta(K_i) =& f_{19} \, P_0 \wedge P_i + \blue{f_{20} \, P_0 \wedge K_i} +\blue{f_{21} \, J_3 \wedge K_i} + \blue{f_{22} \, J_3 \wedge P_i} +
\\
& \left( f_{23} \, P_0 \wedge P_j + \blue{f_{24} \, P_0 \wedge K_j} + \blue{f_{25} \, J_3 \wedge K_j} + \blue{f_{26} \, J_3 \wedge P_j} \right) \varepsilon_{ij} .
\end{split}
\end{equation}
Imposing the cocycle conditions we get the following two independent solutions:
\begin{eqnarray}\label{2+1-spatially_isotropic_sols} 
&&\left\{\begin{aligned}
\delta(P_0) =&   0 \,,\\
\delta(J_3) =&   0 \,, \\
\begin{aligned}
\delta(P_i) =&
\\[5pt]
\\
\delta(K_i) =&
\\[15pt]
\end{aligned}&
\boxed{\begin{aligned}
&    -\Lambda \blue{f_{22}}\red{K_i \wedge J_3}  -\colorboxed{red}{ \Lambda f_{23} \varepsilon_{ij} \red{ K_j\wedge P_0}+ \Lambda f_{23} \red{P_i\wedge J_3}} \\&  -\Lambda \blue{f_{11}} \varepsilon_{ij} \red{ K_j\wedge J_3} +  \blue{f_{11}P_0\wedge P_i} +\varepsilon_{ij} \blue{f_{22}P_0\wedge P_j}
 ,\\
& \colorboxed{red}{ \Lambda  f_{23}\blue{ K_i \wedge J_3} +  f_{23} \varepsilon_{ij}P_0\wedge P_j} - \blue{f_{11}K_i\wedge P_0}\\& -\blue{f_{22}} \varepsilon_{ij} \blue{K_i\wedge P_0}- \blue{f_{22}P_i\wedge J_3}- \blue{f_{11}\varepsilon_{ij}P_j\wedge J_3}.
\end{aligned}}
\end{aligned}\right.
\\
&&\left\{\begin{aligned}
\\[-3pt]
\delta(P_0) &=
\\[7pt]
\delta(J_3) &=
\\[8pt]
\delta(P_i) &= 
\\[8pt]
\delta(K_i) &=
\\[7pt]
\end{aligned}
\boxed{\begin{aligned}
& (\pm f_{23} \sqrt{\Lambda}+\blue{f_{15}})\left(\frac{\Lambda}{2}\varepsilon_{ij}\red{K_i\wedge K_j}\mp\colorboxed{red}{ \sqrt{\Lambda}\varepsilon_{ij}\red{K_i\wedge P_j}}\right)+ (\pm f_{23}\sqrt{\Lambda}+\blue{f_{15}}) \left(\frac{1}{2}\varepsilon_{ij}\blue{P_i\wedge P_j}  \right) \,,\\
&  0 \,, \\
& \colorboxed{red}{ \pm \sqrt{\Lambda} \blue{f_{15}} \red{K_j\wedge P_0} \varepsilon_{ij}+\Lambda f_{23}\red{P_i\wedge J_3}}+\blue{f_{15}P_0\wedge P_j}\varepsilon_{ij}  -\Lambda \blue{f_{15}} \red{K_i\wedge J_3}
,\\
& \pm\sqrt{\Lambda}\varepsilon_{ij}f_{23} \blue{K_j\wedge P_0}\pm\sqrt{\Lambda}f_{23} \blue{P_i\wedge J_3}+\colorboxed{red}{\varepsilon_{ij}f_{23}P_0\wedge P_j\mp \sqrt{\Lambda}\blue{f_{15}K_i\wedge J_3} }
.
\end{aligned}}
\right. \qquad
\label{2+1-spatially_isotropic_sols2}
\end{eqnarray}
All of the terms in the solutions~\eqref{2+1-spatially_isotropic_sols} and \eqref{2+1-spatially_isotropic_sols2} are invariant under $\mP$-transformations (the very ansatz of manifest spatial isotropy forbids non-$\mP$-invariant terms). However not all terms are $\mT$-covariant. 


If in \eqref{2+1-spatially_isotropic_sols} we replace
$\blue{f_{11}} = -z$, $\red{f_{22}}= \red{f_{23}}=0$, we obtain the  $\kappa$-(A)dS Lie bialgebra in 2+1 dimensions~\cite{Angel_q-dS_1994,Ballesteros2017,Ballesteros2017a}:
\begin{equation}
\begin{aligned}
&\delta(P_0) = 0 \,,&  &\delta(P_i) = - z \, P_0\wedge P_i + z \, \Lambda \, \varepsilon_{ij} K_i \wedge J_3 \,,&
\\
&\delta(J_3) = 0 \,,& &\delta(K_i) = - z \, P_0\wedge K_i + + z \, \Lambda \, \varepsilon_{ij} P_i \wedge J_3 \,.&
\end{aligned}
\end{equation}
This Lie bialgebra has been claimed to emerge in the context of (2+1)D Quantum Gravity coupled to point sources~\cite{rosati}. In this model the (topological) gravitational degrees of freedom are under control, and can be integrated away to give rise to an effective kinematics for the point particles, which is deformed by $L_p$-dependent modifications~\cite{matschull}. The spacetime symmetries of this effective model are described by a Hopf algebra.

\subsubsection{$\mP$\&$\mT$ invariance alone}

We can impose $\mP\mT$ invariance without prior imposition of manifest spatial isotropy, and ask whether there are $\mP\mT$-invariant terms that are not manifestly spatially isotropic. 
 As we remarked in Sec.~\ref{math}, although the Poincar\'e algebra is not semisimple, in spacetime dimensions larger than 2 all of its Lie-bialgebra deformations are coboundaries. It is then convenient to discuss $\mP\mT$ invariance at the level of

the $r$-matrix. The following is the most general $\mP\mT$-invariant $r$-matrix:
\begin{equation}
r =  a \, P_1 \wedge P_2 + \red{c \, K_1 \wedge K_2} \,,
\end{equation}
we see that the $r$-matrix is manifestly spatially isotropic, so we fall back into a sub-case of what discussed in Sec.~\ref{SEC_2+1_spatially-isotropic}.

\subsubsection{General case}

Dropping all the assumptions, in the general case too It is convenient to work at the level of the $r$ matrix. Stachura~\cite{Stachura98} classified  all the $r$ matrices on the Poincar\'e algebra, and found five families of solutions, some of which have sub-cases, for a total of eight classes - each consisting of a family of solutions with a different number of free parameters. To attempt at providing some orientation in this multiplicity of solutions, we can distinguish the ones that are more physically relevant because they are the least suppressed by tiny physical constants. These give rise to deformations that are first order in the Planck length and are not suppressed by positive powers of the cosmological constant (the blue terms in Sec.~\ref{DimAn}).
In fact the most general $r$-matrix has the form:
\begin{equation}
r = a + \blue{b} + \red{c}\,, 
\qquad a = \alpha^{\mu\nu} P_\mu \wedge P_\nu \,,
\qquad \blue{b} = \blue{\beta^{\mu\nu\rho} P_\mu \wedge M_{\nu \rho} }\,,
\qquad \red{c} = \red{\gamma^{\mu\nu\rho\sigma} M_{\mu\nu} \wedge M_{\rho\sigma}} \,,
\end{equation}
where the color code corresponds to what each term implies for the cocommutator: 
\begin{equation}
\delta(X) = [r , X \otimes 1 + 1 \otimes X] = [r^{(1)},X] \wedge  r^{(2)} +r^{(1)} \wedge [r^{(2)},X] \,,
\end{equation}
the black terms are of the kind $\mathcal D^{\mu\nu\rho\sigma}$, the blue terms are of the kind $\blue{\mathcal A^{\mu\rho\sigma}}$ and  $\blue{\mathcal E^{\mu\nu\rho\sigma\gamma}}$, while the red terms are of the kind $\red{\mathcal B^{\mu\rho\sigma\gamma}}$ and  $\red{\mathcal F^{\mu\nu\rho\sigma\gamma\delta}}$.

Now, assuming that $a=\red{c}=0$ whike $\blue{b} \neq 0$ excludes two of the classes of solutions described in~\cite{Stachura98}: class I and V. All the other ones correspond to cases in which $\red{c}=0$ and $a$ is generic - and can therefore be put to zero. We are still left with six classes, so our physical characterization of Lie bialgebras is not very discriminatory. We will show below that the same requirements are much more stringent in 3+1 dimensions.

\subsection{Four spacetime Dimensions}

\subsubsection{Manifest spacetime covariance}

A basis of invariant tensors for the coefficients must be written in terms of  $\varepsilon_{\mu\nu\rho\sigma}$ and the metric, which in (3+1)D are both even-rank. So  only the even-rank coefficients $\red{\mathcal{B}^{\mu\rho\sigma\gamma}}$, $\mathcal{D}^{\mu\nu\rho\sigma}$ and $\red{\mathcal{F}^{\mu\nu\rho\sigma\gamma\delta}}$ can be written as invariant tensors. To our knowledge no minimal tensor basis classification has been made in this case, so we generate all possible
terms\footnote{By doing this we generate an overcomplete basis, which is not a problem because it just means that we write each independent numerical coefficient as a linear combination of the coefficients of our overcomplete basis.} made with $\varepsilon_{\mu\nu\rho\sigma}$ and $\eta_{\mu\nu}$. 

Imposing the cocycle condition and co-Jacobi identities on the cocommutator written this way gives no nontrivial solution:
\begin{equation*}\label{eq:5} 
\delta(P_\mu) = 0, \qquad
\delta(M_{\mu\nu}) = 0.
\end{equation*}
Hence there is no spacetime-isotropic Lie bialgebra deformation of the Poincar\'e or (A-)dS algebra in 3+1 dimensions.

\subsubsection{Manifest spatial isotropy}\label{4D_spatial_isotropy_sec}

The most general spatially isotropic cocommutator in the $10$-dimensional basis $\{P_0, J_i, P_i, K_i  \}$ is
\begin{equation*}\label{eq:23} 
\begin{split}
\delta(P_0) =& \red{f_1 \, P_i \wedge J^i} + \red{f_2 \, P_i \wedge K^i}+\red{f_3 \, J_i \wedge K^i}, \\
\delta(J_i) =& P_0 \wedge (\blue{f_4 \, J_i} + f_5 \, P_i + \blue{f_6 \,  K_i}) +f_7 \, P_j \wedge P_k \varepsilon_{ijk}+
\\ 
&( \red{f_8 \,J_j \wedge J_k} + \red{f_9 \,K_j \wedge K_k} + \blue{f_{10} \,P_j \wedge J_k} + \blue{f_{11} \,P_j \wedge K_k} +\red{f_{12} \, J_j \wedge K_k})\varepsilon_{ijk}, \\
\delta(P_i) =& P_0 \wedge (\red{f_{13} \,J_i} + \blue{f_{14} \,P_i} + \red{f_{15} \,K_i}) +\blue{f_{16} \,P_j \wedge P_k}\varepsilon_{ijk} + 
\\&(\red{f_{17} \,J_j \wedge J_k}+\red{f_{18} \, K_j \wedge K_k} +\red{f_{19} \, P_j \wedge J_k} + \red{f_{20} \,P_j \wedge K_k} + \red{f_{21} \, J_j \wedge K_k})\varepsilon_{ijk}, \\
\delta(K_i) =& P_0 \wedge (\blue{f_{22} \,J_i}+f_{23} \, P_i+\blue{f_{24} \,K_i}) +f_{25} \, P_j \wedge P_k \varepsilon_{ijk}+ \\
&(\red{f_{26} \,J_j \wedge J_k}+ \red{f_{27} \,K_j \wedge K_k} + \blue{f_{28} \,P_j \wedge J_k} + \blue{f_{29} \,P_j \wedge K_k}+\red{f_{30} \, J_j \wedge K_k})\varepsilon_{ijk}.
\end{split}
\end{equation*}
Imposing the Lie bialgebra axioms one obtains the unique  solution:
\begin{equation}\label{kappa-Poincare} 
\begin{split}
\delta(P_0) &= 0, \\
\delta(J_i) &= 0, \\
\delta(P_i) &= \boxed{( 1 - |\lambda|) \blue{f_{14} \, P_0\wedge P_i}}, \\
\delta(K_i) &= \boxed{( 1 - |\lambda|)  \blue{f_{14}} \left( \blue{P_0 \wedge K_i} +\blue{ \varepsilon_{ijk}P_j\wedge J_k } \right) }.
\end{split}
\end{equation}
which is the (timelike\footnote{The $\kappa$-Poincar\'e $r$-matrix $r= {\frac 1 \kappa} K_i\wedge P_i$ is actually a particular choice among a $4$-parameter family of solutions, parametrized by a vector $v_\mu \in \mathbb{R}^4$ (see below). The most studied version of the algebra is characterized by a timelike vector $v^\mu = \frac 1 \kappa \, \delta^\mu{}_0$ (the energy scale $\kappa$ gives the name to the algebra). It is still not entirely clear whether different choices of vector correspond to nonequivalent physics. }) $\kappa$-Poincar\'e Lie bialgebra.\footnote{For an explicit ``exponentiation" of the $\kappa$-Poincar\'e Lie bialgebra to obtain the $\kappa$-Poincar\'e Hopf algebra see for instance \cite{ang}.} The cocommutator above is covariant under $\mP$, but not under $\mT$ nor $\mP \mT$.

This is  a significant and original result which shows that in $3+1$ spacetime dimensions only $\kappa$-Poincar\'e  and $\kappa$-Minkowski can serve as manifestly isotropic quantum algebra and quantum spacetime, not only considering $\mathfrak{iso}(3, 1)$ but the  $\text{(A-)dS}$ cases too.

\subsubsection{$\mP$\&$\mT$ invariance alone}

Imposing, as above, the separate $\mP$- and $\mT$-invariance of the $r$-matrix, we get the following:
\begin{equation}
\begin{aligned}
r =& \red{f_1 \, K_1 \wedge K_2 + f_2 \, K_1 \wedge K_3 + f_3 \, K_2 \wedge K_3} + f_4 \, P_1 \wedge P_2 + f_5 \,P_1 \wedge P_3 + f_6 \, P_2 \wedge P_3 + \\
&\red{f_7 \, R_1 \wedge R_2 + f_8 \, R_1 \wedge R_3 + f_9 \, R_2 \wedge R_3} \,.
\end{aligned}
\end{equation}
We see that in 4D the most general $\mP$- and $\mT$-invariant $r$-matrix is not necessarily manifestly space-isotropic.

Imposing the co-Jacobi conditions we get three independent nontrivial solutions in the (A)dS case $\lambda \neq 0$:
\begin{equation}
\red{f_7}^2+\red{f_8}^2+\red{f_9}^2=0 \,, 
\qquad \left\{
\begin{aligned}
&\red{f_1}=-\red{f_7} \,, ~ \red{f_2}=-\red{f_8} \,, ~ \red{f_3}=-f_ 9\,~  f_4=f_5=f_6=0\,,
\\
&\red{f_1}=\red{f_2}=\red{f_3}=0 \,,~  \Lambda f_4 =  \red{f_7} \,, ~ \Lambda f_5 = \red{f_8} \,, ~ \Lambda f_6 = f_ 9\,, 
\\
&\red{f_1}=\red{f_2}=\red{f_3}=f_4=f_5=f_6=0 \,,
\end{aligned}
\right.
\end{equation}
the first condition, common to all of the three solutions, admits no real nonzero solution.

In the Poincar\'e case $\Lambda=0$ we have the $\Lambda \to 0$ limit of the three solutions above, plus this additional solution:
\begin{equation}
\red{f_1}=\red{f_2}=\red{f_3}=\red{f_7}=\red{f_8}=\red{f_9}=0 \,,
\end{equation}
that is, $f_4$, $f_5$ and $f_6$ are the only nonzero parameters, and they are freely specifiable. This solution is then:
\begin{equation} 
\colorboxed{red}{\boxed{\delta(P_\mu) = 0\,, ~~~
\begin{aligned}
\delta(K_1) =& f_4 \, P_0 \wedge P_2 + f_5 \, P_0  \wedge  P_3\,,\\
\delta(K_2) =& f_6 \, P_0 \wedge P_3 - f_4 \, P_0  \wedge  P_1\,,\\
\delta(K_3) =& -f_5 \, P_0 \wedge P_1 - f_6 \, P_0  \wedge  P_2\,,
\end{aligned}
~~~
\begin{aligned}
\delta(J_1) =& f_4 \, P_1 \wedge P_3 - f_5 \, P_1  \wedge  P_2\,,\\
\delta(J_2) =& f_4 \, P_2 \wedge P_3 - f_6 \, P_1  \wedge  P_2\,,\\
\delta(J_3) =& f_5 \, P_2 \wedge P_2 - f_6 \, P_1  \wedge  P_3\,.\\
\end{aligned}}}
\end{equation}
This is the only example of $\mP$- and $\mT$-invariant Lie bialgebra based on the Poincar\'e algebra in 3+1 dimensions. It cannot be obtained as a contraction of an (A)dS analogue, and its three independent coefficients have the dimensions of a squared length. This case deserves further investigation.

\subsubsection{General $\Lambda=0$ case and a no-go theorem}\label{gen}

Just like in the 2+1-dimensional case, the most generic Lie bialgebra
over the 3+1D Poincar\'e algebra is a coboundary, and the $r$ matrices have in this case too been classified. In this case the relevant reference is a paper by Zakrzewski~\cite{zaz}, in which the author finds 23 classes of $r$ matrices that solve the modified Yang--Baxter equation. Just like in 2+1D, the $r$-matrix can be decomposed as $r = a + \blue{b} + \red{c}$, and the most `physically interesting' terms are the blue ones, that is, terms of type $ \blue{b}$. Imposing that $ a = \red{c} =0$ results in a remarkable simplification: there is just a 7-parameter family of solutions, given by
\begin{equation}\label{zzza}
r =  \blue{v^\mu \, M_{\mu\nu} \wedge P^\nu + v^\mu P_\mu \wedge  M(u_1,u_2,u_3)\,,}
\end{equation}
where $v^\mu$ is a real 4-vector and $M(u_1,u_2,u_3)$ is a generic element of the stabilizer, in $\mathfrak{so}(3,1)$, of $v^\mu P_\mu$. The latter depends on three real parameters $u_1$, $u_2$ and $u_3$, because the stabilizer of a 4-vector is a 3-dimensional subalgebra: $\mathfrak{so}(3)$ in case $v^\mu$ is timelike, $\mathfrak{so}(2,1)$ if it is spacelike and $\mathfrak{iso}(2)$ if it is lightlike. If $u_1-u_2=u_3=0$ we obtain the
%
%
%
%
the vector-like generalization of the $\kappa$-Poincar\'e algebra\cite{koso}:
%
%
\begin{equation}\label{kappa-Poincare_vectorlike} 
\boxed{\begin{split}
\delta(P_\mu) &= \blue{v^\nu \, P_\nu \wedge P_\mu } \,, \\
\delta(M_{\mu\nu}) &=     \blue{v_\nu \, M_{\rho\mu} \wedge P^\rho}  - \blue{v_\mu \, M_{\rho\nu} \wedge P^\rho} \,.
\end{split}}
\end{equation}
 When any of the $u_i$ parameters are nonzero, the additional term in the $r$ matrix generates a \emph{Reshetikhin twist}~\cite{TwistedKappa0,TwistedKappa1,TwistedKappa2,TwistedKappa3}, 
which changes significantly the cocommutators. For example, in the case in which $v^\mu$ points in the time direction, $v^\mu = {\frac 1 \kappa} \delta^\mu{}_0$, the $r$-matrix can be written 
\begin{equation}\label{zzza2}
r =  \blue{ \frac 1 \kappa \left( P_j \wedge K^i +  P_0 \wedge  u^i J_i\right) \,,}
\end{equation}
which generates the following cocommutator:
\begin{equation}\label{kappa-Poincare_vectorlike_twisted} 
\boxed{\begin{aligned}
\delta(P_0) &= 0\,, \\
\delta(P_i) &= \blue{\frac 1 \kappa \left( P_0 \wedge P_i + \varepsilon_{ijk} u_j \, P_0 \wedge P_k \right)} \,, \\
\delta(R_i) &= \blue{\frac 1 \kappa \left( \varepsilon_{ijk} u_j \, P_0 \wedge R_k  \right)}\,, \\
\delta(K_i) &= \blue{\frac 1 \kappa \left( P_0 \wedge K_i + \varepsilon_{ijk} u_j \, P_0 \wedge K_k + u_j \, P_i \wedge R_j  - \varepsilon_{ijk} P_j \wedge R_k \right) }\,. \\
\end{aligned}}
\end{equation}
In summary: our work proves that  assuming manifest spatial isotropy alone, the $\kappa$-deformation is the only one extending relativistic symmetries to a  4-dimensional noncommutative setting. Zakrzewski's analysis \textit{plus} our  dimensional/regularity considerations yields the full 7-parameter family of \emph{twisted vector-like generalized $\kappa$-deformations}~\cite{TwistedKappa0,TwistedKappa1,TwistedKappa2,TwistedKappa3}
.\footnote{Of this 7-parameter family, if one considers Lie-bialgebra automorphisms, only three cases are truly independent: when the $v^\mu$ vector is space-, time- or light-like~\cite{nullplane,spacelike}.}
This results  can be seen as a no-go theorem regarding the existence of physically-meaningful Lie-bialgebra deformations other than (twisted) $\kappa$-Poincar\'e in 4D.

\subsubsection{Fully general $\Lambda \neq 0$ case}

If the cosmological constant is not zero, Zakrzewski's analysis does not apply. We have to write the most general $r$-matrix, and even imposing the dimensional/regularity assumption of keeping only `blue' terms does not lead to a treatable set of equations. We thus postpone the investigation of this case in full generality to future works.

\section{Conclusions}\label{conc}

In this paper we addressed the issue of finding quantum deformations of relativistic symmetry Lie algebras: Poincar\'e, de Sitter and Anti-de Sitter.
Our analysis was carried through a systematic, algorithmic approach. We classified our results through dimensional analysis, analytic flat limit and manifest covariance principles.

We get two main results, which can be understood as no-go theorems for alternatives to $\kappa$-Poincar\'e.
The first result is that, under the assumption of manifest spatial isotropy, the unique deformation which generalizes the relativistic Poincar\'e algebra in (3+1)D, is the so-called $\kappa$-Poincar\'e deformation, whose homogeneous spacetime is $\kappa$-Minkowski \cite{lukkk,majid22,fla}. 
The second result is that, requiring the Lie-bialgebra deformation to depend on structure constants that 1. are first-order in the Planck length $L_p$ and 2. admit a well-defined flat ($\Lambda \to 0$) limit, the only deformation of the Poincar\'e algebra in (3+1)D is the `twisted vector generalization' of the $\kappa$-Poincar\'e Lie bialgebra~\cite{TwistedKappa0,TwistedKappa1,TwistedKappa2,TwistedKappa3}. 
 This is a double modification of the original, so-called `timelike'  $\kappa$-Poincar\'e Lie bialgebra: a four-parameter family of terms can be obtained from the original Lie bialgebra by taking a linear combination of the generators that `rotates' the primitive (zero-cocommutator) generator from $P_0$ to $v^\mu P_\mu$, where $v^\mu$ are four arbitrary parameters. The rest is a 3-parameter perturbation which is generated by a Reshetikhin twist~\cite{TwistedKappa0,TwistedKappa1,TwistedKappa2,TwistedKappa3},
 which depends on three real numbers $u_i$ - the parametrization of the stabilizer of a 4-vector in $\mathfrak{so}(3,1)$.

These no-go theorems point out that, if we want to work at first order in the Planck length, the only viable option is (possibly twisted) $\kappa$-Poincar\'e and its possible generalizations into (Anti-)de Sitter. To elude this result, we have to consider deformations that depend on the square of the Planck length (an awfully small quantity, which makes testable predictions even harder to find than they already are).
Further insights regarding generalizations of the $\kappa$-Poincar\'e algebra can be gained by comparing our results with those in~\cite{ballesteroscurved}. In that work the authors identify the minimal assumptions that one should make in order to uniquely characterize the  $\kappa$-(Anti) de Sitter Lie bialgebra. Of course spatial isotropy is not among these assumptions, otherwise, as we have shown in~\ref{4D_spatial_isotropy_sec}, one is left only with $\kappa$-Poincar\'e in the $\Lambda =0$ case and nothing otherwise. As a result, the authors of~\cite{ballesteroscurved}  find that $\Lambda \neq 0$ implies a non-trivial Planck-scale deformation of rotations.

In (2+1)D, there appears to be a richer catalog of Hopf algebraic structures, generalizing Poincar\'e and (A-)dS invariance. We believe that these frameworks represent the starting point for further investigations in (2+1)D Quantum Gravity. Although we don't have a full classification of these deformations (there are hundreds of solutions of the Lie-bialgebra axioms), in this paper we provided a first classification of the most physically-interesting ones, based on dimensional analysis, various degrees of manifest isotropy and discrete symmetries.

Finally, the analysis of the cases in which discrete symmetries like $\mP$ and $\mT$ invariance are implemented at the level of the Lie bialgebra led to a new result in (3+1)D. In the flat, $\Lambda = 0$ case, there is a 3-parameter family of Lie-bialgebra deformations of the Poincar\'e algebra which implements these discrete symmetries as Lie-bialgebra automorphisms. These deformations depend quadratically on the Planck length and therefore elude the no-go theorems found above. They represent therefore an interesting starting point to begin the exploration of these `order $L_p^2$' Lie bialgebras.

\section*{Acknowledgements}

We would like to thank \'Angel Ballesteros and Giulia Gubitosi for their useful comments and remarks.  F.M. was funded by the Euroean Union and the Istituto Italiano di Alta Matematica
under a Marie Curie COFUND action.

\vspace{12pt}

\providecommand{\href}[2]{#2}\begingroup\raggedright\endgroup

\end{document}